\documentclass[conference]{IEEEtran}
\IEEEoverridecommandlockouts
\usepackage[utf8]{inputenc} 
\usepackage[T1]{fontenc}    
\usepackage{hyperref}       
\usepackage{url}            
\usepackage{booktabs}       
\usepackage{amsfonts}       
\usepackage{nicefrac}       
\usepackage{microtype}      
\usepackage{lipsum}		
\usepackage{graphicx}
\usepackage{doi}
\usepackage{balance}

\usepackage{graphicx}
\usepackage{float}
\usepackage{amsmath}
\usepackage{setspace} 
\usepackage{units}
\usepackage{bmpsize}
\usepackage{array}
\usepackage{empheq}
\usepackage{algorithm}
\usepackage{algpseudocode}
\usepackage{enumitem}   
\usepackage[caption=false]{subfig}

\newtheorem{thm}{Theorem}[section]

\newtheorem{rem}[thm]{Remark}

\title{
Nonsmooth-Optimization-Based Bandwidth Optimal Control for Precision Motion Systems}

\author{
\IEEEauthorblockN{Jingjie Wu}
\IEEEauthorblockA{Department of Mechanical Engineering\\
	The University of Wisconsin-Madison\\
	Madison, WI, 53706 \\
	\texttt{jingjie.wu@wisc.edu} }
\and
\IEEEauthorblockN{Lei Zhou}
\IEEEauthorblockA{Department of Mechanical Engineering\\
	The University of Wisconsin-Madison\\
	Madison, WI, 53706 \\
	\texttt{lei.zhou@wisc.edu} }

}

\begin{document}

\maketitle
\thispagestyle{empty}
\pagestyle{empty}

\begin{abstract}
Precision motion systems are at the core of various manufacturing equipment. The rapidly increasing demand for higher productivity necessitates higher control bandwidth in the motion systems to effectively reject disturbances while maintaining excellent positioning accuracy. However, most existing optimal control methods do not explicitly optimize for control bandwidth, and the classic loop-shaping method suffers from conservative designs and fails to address cross-couplings, which motivates the development of new control solutions for bandwidth optimization. 
This paper proposes a novel bandwidth optimal control formulation based on nonsmooth optimization for precision motion systems. Our proposed method explicitly optimizes the system's MIMO control bandwidth while constraining the H-infinity norm of the closed-loop sensitivity function for robustness. A nonsmooth optimization solver, GRANSO, is used to solve the proposed program, and an augmented quadratic programming (QP)--based descent direction search is proposed to facilitate convergence. Simulation evaluations show that the bandwidth optimal control method can achieve a 23\% higher control bandwidth than conventional loop-shaping design, and the QP-based descent direction search can reduce iteration number by 60\%, which illustrates the effectiveness and efficiency of the proposed approach.

\end{abstract}

\section{Introduction}\label{sec:intro}


Precision motion systems play a critical role in a wide range of manufacturing applications such as photolithography~\cite{oomen2018advanced}, electronic packaging~\cite{ding2006motion}, and wafer inspection~\cite{albero2011micromachined}. With the demand for manufacturing throughput drastically increasing, the motion systems in the manufacturing machines are required to achieve higher speed, which shifts the disturbance signals (including speed-dependant system nonlinearity, motor force/torque ripple, and payload variation) to a higher frequency range. To effectively reject these disturbances during high-speed motion, the feedback control system must achieve higher control bandwidth. Although several optimal control methods (such as mixed sensitivity $H_\infty$ control \cite{ortega2004systematic}) are tailored for frequency-domain design specifications, they often do not explicitly optimize for control bandwidth, and thus are often unsuitable for controller synthesis for high-bandwidth precision motion systems. The classic loop-shaping technique \cite{aastrom2006advanced} effectively designs for frequency-domain performances and still serves as the most widely used controller synthesis tool in the industry; however, it does not guarantee optimality in control bandwidth and often results in conservative controllers, especially for systems with multiple degrees of freedom (DOFs) exhibiting cross-couplings. This fact motivates the development of new and effective optimal control solutions that can explicitly optimize for control bandwidth while satisfying robustness criteria and other design constraints to meet the rapidly increasing needs. 


In the past decade, aiming to address the aforementioned challenge, several controller synthesis methods for improving control bandwidth have been investigated. For example, Van der Veen et al.~\cite{van2015integrated, van2017integrating} proposed an integrated topology and controller optimization framework for motion stages, where the design objective aims to minimize the magnitude of the sensitivity function at a selected low-frequency point, thereby enhancing the control bandwidth. 
Dumanli et al. \cite{dumanli2018optimal} considered a ball-screw drive with acceleration and jerk feedback and proposed a pole-placement method with objective weights tuning to optimize for control bandwidth. Although effective in certain case studies, the enhancement of control bandwidth is achieved indirectly in these methods by considering other intuitive measures, which cannot guarantee bandwidth optimality. 
Another group of efforts uses controller parameters to approximate the control bandwidth. For example, 
Ding et al. \cite{ding2020optimal} and Delissen et al. \cite{delissen2023integrated} assume a free-mass plant for the motion system and approximate the control bandwidth via a controller parameter. This method demonstrated good performance for systems with the assumed dynamics, but can fail when more complicated dynamics (e.g., structural resonances and coupling effects) exist in the system dynamics. 
To directly optimize for control bandwidth, Wu et al. \cite{wu2022control} presented a control co-design framework where mixed-sensitivity $H_\infty$ control is used for controller synthesis, and the bandwidth is directly optimized by searching the optimal weighting filter parameters. In this formulation, the objective function explicitly includes the control bandwidth, and the cross-coupling effects are considered. However, the direct-search-based bandwidth optimization together with mixed-sensitivity $H_\infty$ control is time-consuming, which limits its applicability for practical controller synthesis for motion systems. 

Aiming to provide an efficient and high-performance controller synthesis tool for precision motion systems, this paper proposes a novel bandwidth optimal control formulation based on nonsmooth optimization. The proposed framework directly takes the multi-input multi-output (MIMO) closed-loop control bandwidth as the objective function, and robustness is considered by constraining the $H_{\infty}$ norm of the MIMO sensitivity function. Nonsmooth optimization is used to address the challenge that the objective function and constraints are not continuously differentiable at some points, and a steepest descent direction calculation method based on quadratic programming (QP) is introduced for the nonsmooth optimization to facilitate convergence. The proposed method is tested to synthesize a controller for a high-performance magnetically levitated precision positioning system called FleXstage \cite{JingjieASPE_2022}. The resultant controller can achieve a high control bandwidth of 123~Hz, which is 23\% higher than the conventional loop-shaping design. In addition, simulations show that the inclusion of the QP-based descent direction search can reduce the number of iterations by 60\% and thus effectively enhance the computational efficiency. 

The rest of this paper is organized as follows. Section~\ref{sec:statement} introduces the problem statement. Section~\ref{sec:formulation} presents the proposed bandwidth optimal control problem formulation. Section~\ref{sec:nonsmooth} illustrates the nonsmoothness in the optimization problem and discusses its solving. Section~\ref{sec:simulations} presents the simulation evaluations. Conclusion and future work are discussed in Section~\ref{sec:conclusion}.


\section{Problem Statement}\label{sec:statement}
The dynamics of a motion system considering its flexible dynamics can typically be written as
\begin{align} \label{eqn:mech_EOM}
\begin{split}
M\ddot{x}+D\dot{x}+Kx &= Pu,\\
    y &= Qx,
\end{split}
\end{align}  
where $x$ is the state variable vector including both the rigid-body displacements and flexible modal displacements; $M$, $D$, $K$ are diagonal mass, damping, and stiffness matrices,  respectively, $P$ is the input matrix, $Q$ is the measurement matrix, $u$ is the input vector, and $y$ is the measurement vector.

The \textit{bandwidth optimal control problem} for system~\eqref{eqn:mech_EOM} can be roughly formulated as: synthesize a feedback controller that \textit{maximizes the system's closed-loop control bandwidth} while satisfying robustness constraints, providing stability guarantee, and satisfying other design constraints (e.g., gain limits and control output limits). 

\section{Bandwidth Optimal Control Problem Formulation}\label{sec:formulation}
This section introduces the formulation of the bandwidth optimal control problem. As a first step, the dynamics of the precision motion systems is typically transformed into $n$ decoupled channels. Here, transformations $\hat{u} = T_u u$ and $\hat{y}=T_y y$ 
are applied to the dynamics in \eqref{eqn:mech_EOM} and obtain  
\begin{align} \label{eqn:mech_EOM_decp}
\begin{split}
M\ddot{x}+D\dot{x}+Kx &= \hat{P}\hat{u},\\
    \hat{y} &= \hat{Q}x,
\end{split}
\end{align} 
where $\hat{P}=PT_u^{-1}$ and $\hat{Q}=T_y Q$ are the decoupled input and output matrices with almost diagonal structures in first $n\times n$ components, respectively; $\hat{u} \in \mathbb{R}^n$ is the recoupled force input, $\hat{y} \in \mathbb{R}^n$ is the decoupled measurement output, and $n$ is the number of total decoupled DOFs to be controlled. The decoupled system plant transfer function $G(s)$ can then be derived as 
\begin{align} \label{eqn:G_decp}
\hspace{-5pt}
\setlength{\arraycolsep}{0.5pt}
    G(s): \begin{bmatrix}  \dot{x} \\ \ddot{x} \\ \hat{y} \end{bmatrix} = 
    \begin{bmatrix}
        0  & I & 0 \\
        -M^{-1} K  & -M^{-1} K & M^{-1} \hat{P} \\
        \hat{Q} & 0 & 0
    \end{bmatrix} 
    \begin{bmatrix}
        x\\ \dot{x} \\ \hat{u}
    \end{bmatrix},
\end{align}
where $s$ is the Laplace variable. 

\begin{figure}[t!]
\centering
\subfloat{
\includegraphics[trim={0mm 0mm 0mm 0mm},clip,width =1\columnwidth, keepaspectratio=true]{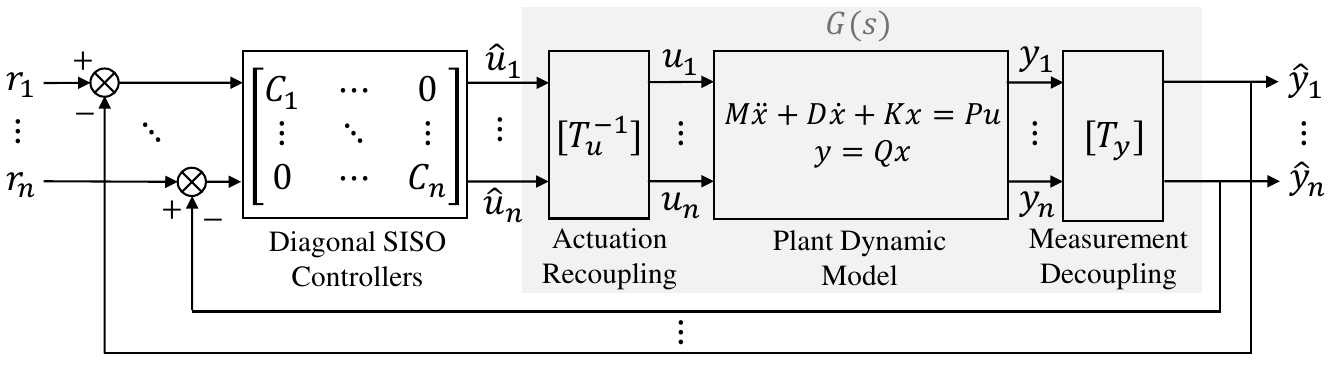}}
\vspace{-4mm}
\caption{Control block diagram of a typical decoupled motion system. $T_u$ is the force decoupling matrix and $T_y$ is the measurement decoupling matrix. The resultant decoupled system model $G$ is mainly diagonal with cross-coupling terms on off-diagonal entries. }
\label{fig:control_diagram}
\vspace{-3mm}
\end{figure}

For most motion systems, $n$ typically equals the number of motion axes. For some systems with extra flexible modes controlled by over-actuation \cite{wu2022sequential}, $n$ includes the number of rigid-body motion axes and the number of flexible modes under feedback control. For decoupled multi-axes plant dynamics \eqref{eqn:G_decp}, decentralized controllers are typically used, where a single-input, single-output (SISO) controller is used for each decoupled channel. Figure~\ref{fig:control_diagram} illustrates the decoupling of the system. 


The bandwidth optimal control problem for a decoupled multi-axes motion system \eqref{eqn:G_decp} is formulated as 
\begin{align} \label{eqn:opt_form}
\begin{split}
    \max_{\theta_c \in \mathbb{R}^m} ~~~&\omega_{bw},
\\
    \mathrm{s.t.}~~~&\| S(\theta_c) \| _{\infty} \leq S_{max}, 
\\
                    &\gamma(\theta_c) \leq 0,
\end{split}
\end{align}
where $S=(I+GC)^{-1}$ is the closed-loop sensitivity function, $C$ is the MIMO feedback controller,  $\theta_c \in \mathbb{R}^m$ is a vector for the parameters of controller $C$, $\omega_{bw}$ is the system bandwidth, and $\gamma(\theta_c)$ indicates other control system design constraints.  $\| S\|_{\infty}$ is the $H_{\infty}$-norm of $S$ computed as
\begin{align} \label{eqn:Hinfnorm}
    \| S(s)\|_{\infty} = \sup_{\omega \in \mathbb{R}} \Bar{\sigma}(S(j\omega)),
\end{align}
where $\Bar{\sigma}$ is the maximum singular value, and $S_{max}$ is the constraint value for $\| S\|_{\infty}$.

The definition of bandwidth in \eqref{eqn:opt_form} can be selected according to the system's needs. In this work, we define the bandwidth as the system's cross-over frequency, i.e., the first frequency satisfying
\begin{align} \label{eqn:bw_def}
    \underline{\sigma}(L(j\omega)) = 1,
\end{align}
where $L$ is the loop gain of the largely-decoupled MIMO system computed as $L =GC$, and $\underline{\sigma}$ represents the minimum singular value of a matrix. 
This selection is suitable for motion systems since their plants and loop gains are typically low-pass in nature. It also unifies the bandwidth performance of all SISO channels and thus simplifies the optimization formulation and computation. 
Of note, the problem \eqref{eqn:opt_form} considers the closed-loop stability and robustness of the overall MIMO system instead of individual decoupled SISO channels, thereby considering the cross-coupling between motion axes.

\rem{It is worth noting that \eqref{eqn:bw_def} serves as an implicit constraint defining the objective function $\omega_{bw}$.  In addition, the stability guarantee of the bandwidth optimal control problem \eqref{eqn:opt_form} is implicitly included by the robustness constraint, since a linear-time invariant system is Lyapunov stable if and only if (iff)  $\| S(s)\|_{\infty}$ is finite~\cite{desoer2009feedback}.}

The form of the SISO controllers on the diagonal of $C$ is determined by the system being controlled. 
In this work, we select $C = \mathrm{diag}\{C_1, C_2, ..., C_n\}$, where $C_i, i = 1, \dots, n$ takes a commonly-used PID controller parameterization as~\cite{butler2011position} 

\begin{small}
  \vspace{-6pt}
\begin{align}\label{eqn:PID}
\medmuskip=-4mu
\thinmuskip=-4mu
\thickmuskip=-4mu
    C_i(s) = K_p\Big(\frac{s+\omega_I}{s}\Big)\Big(\frac{s}{\omega_D}+1\Big)
    \Big(\frac{1}{\frac{s^2}{\omega_{lp}^2}+\frac{2z_{lp}s}{\omega_{lp}}+1}\Big).
\end{align}
\end{small}

\noindent The controller \eqref{eqn:PID} is a PID controller with a second-order low-pass filter, and the definition of parameters is described in Table~\ref{table:PID_para}. Note that with other design parameters depending on $\omega_c$, $\alpha$, and $m$ and with  $\alpha$ and $m$ fixed for a certain system, the controller $C_i$ solely depends on one single parameter $\omega_c$. This controller form is often used for motion systems including references \cite{heyman2023levcube, steinbuch1998advanced}. 

\begin{table}[t] 
\centering
\vspace{4mm}
\caption{PID controller parameters for \eqref{eqn:PID} \cite{butler2011position}. }
\label{table:PID_para}
    \begin{center} \begin{small}
        \begin{tabular}{ p{1.2cm} p{4.cm}  p{1.5cm} }
        \hline
        Parameter & Description &  Value \\
        \hline
        $\omega_c$  & Desired bandwidth [rad/s]  & --  \\
        
        $m$    &  Modal mass    & --  \\
        
        
        $\alpha$ & PID frequency ratio  &   3 \\
        
        $K_p$  &    Proportional gain  & $m \omega_c^2/\alpha$   \\
        
        $\omega_I$  & Integrator frequency  & $  \omega_c/ \alpha^2$  \\
        
        $\omega_D$  & Differentiator frequency  & $\omega_c/\alpha$ \\
        
        $\omega_{lp}$  & Low-pass filter frequency &  $  \alpha \omega_c$  \\
        
        $z_{lp}$  & Low-pass filter damping ratio  & 0.7\\
        \hline

        \end{tabular}
    \end{small} \end{center}
\vspace{-8mm}
\end{table}

The PID controller \eqref{eqn:PID} effectively shapes the loop for typical rigid motion systems. However, for systems with mode couplings and structural resonances, the controller  \eqref{eqn:PID} suffers from low design freedom and can lead to conservative performance. To address this, notch filters in the following form are typically used in addition to the PID controller:
\begin{align} \label{eqn:notch}
    N(\beta,\zeta,s)=\frac{s^2+2\beta\zeta \omega_n s + \omega_n^2}{s^2+2\zeta \omega_n s + \omega_n^2},
\end{align}
where $\omega_n$ is the notch frequency, and $\beta$ and $\zeta$ represent the depth and width of the notch, respectively. 
For a motion with $n$ decoupled axes, $n$ PID controllers are used, and $p$ notch filters are included in all axes. The 
vector of controller parameters is $\theta_c = [\boldsymbol{\omega}_{\mathbf{c}}^\top, \boldsymbol{\beta}^\top, \boldsymbol{\zeta}^\top]^\top$, where $\boldsymbol{\omega_{\mathbf{c}}} = [\omega_{c1}, ..., \omega_{cn}]^\top$, $\boldsymbol{\beta}=[\beta_{1}, ..., \beta_{p}]^\top$, and $\boldsymbol{\zeta}= [\zeta_{1}, ..., \zeta_{p}]^\top$. The number of decision variables of problem \eqref{eqn:opt_form} for this system is $m = n+2p$.

\section{Nonsmooth Optimization for Bandwidth Optimal Control}\label{sec:nonsmooth}

\subsection{Nonsmoothness in Bandwidth Optimal Control}

We make a key observation that both the objective function and robustness constraint in the bandwidth optimal control problem \eqref{eqn:opt_form} are not continuously differentiable at some local critical points, i.e., exhibiting nonsmoothness. This property incurs challenges in solving the problem, and regular gradient-based solvers are unsuitable. This section discusses the nonsmoothness in the bandwidth optimal control problem. 

\subsubsection{Nonsmoothness in Objective} 
\begin{figure}[t!] 
\centering
\subfloat{
\includegraphics[trim={0mm 0mm 0mm 0mm},clip,width =0.95\columnwidth, keepaspectratio=true]{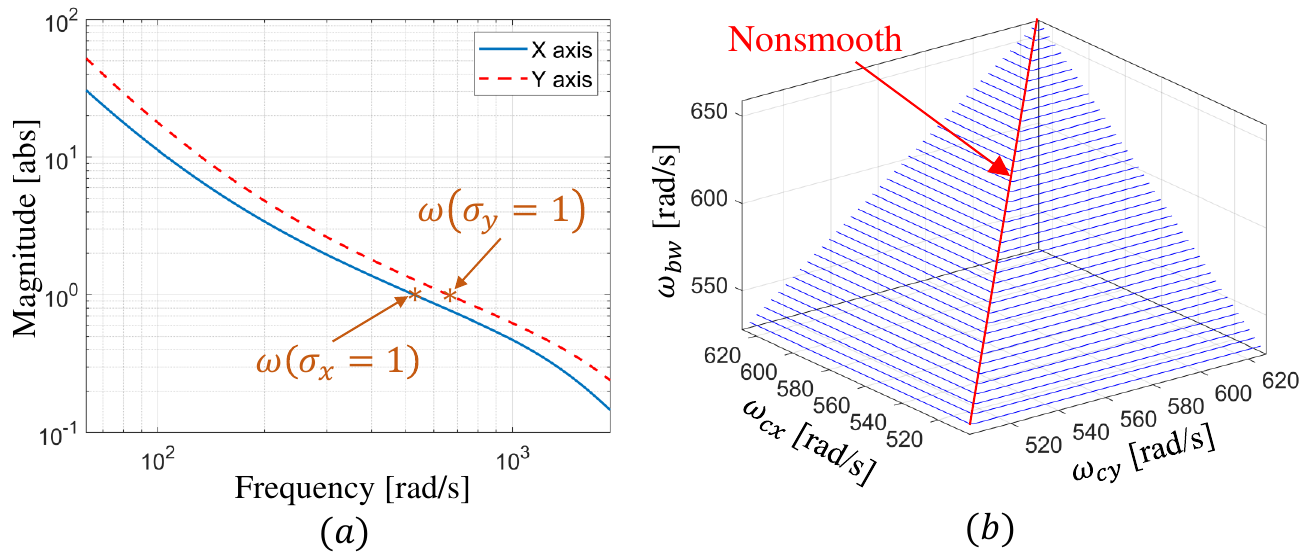}}
\vspace{-4mm}
\caption{(a) Loop Gains of X and Y axis motion. (b) Contour plot of $\omega_{bw}$ values in terms of different $\omega_{cx}$ and $\omega_{cy}$.}
\label{fig:obj_nonsmooth}
\vspace{-3mm}
\end{figure}

The nonsmoothness in the objective function $\omega_{bw}$ 
occurs when the minimum singular value of $L(j\omega_{bw})$ has a multiplicity larger than 1. 
Consider an example $x$-$y$ motion system with two input forces and two displacement feedback signals. A SISO controller as \eqref{eqn:PID} is used for each axis, and $\omega_{cx}$ and $\omega_{cy}$ are the controller parameters for the corresponding axes. Fig.~\ref{fig:obj_nonsmooth}a illustrates the frequency responses of the $x$- and $y$-axis loop gain with the frequencies satisfying $\sigma(L(j\omega))=1$ highlighted, where $\sigma(\cdot)$ represents the singular value. When $\omega_{cx} = \omega_{cy}$, the two singular values of $L$ are repeated at $\omega_{bw}$ and simultaneously being the minimum singular value of $L$. 
Figure~\ref{fig:obj_nonsmooth}b illustrates the contour plot of $\omega_{bw}$ with respect to $\omega_{cx}$ and $\omega_{cy}$. It can be observed that $\omega_{bw}$ is not differentiable along a ridge $\omega_{cx}=\omega_{cy}$. When at the nonsmooth location, varying a single parameter cannot increase $\omega_{bw}$, and the steepest ascent happens along the ridge. 


\begin{figure}[t!]
\centering
\subfloat{
\includegraphics[trim={0mm 0mm 0mm 0mm},clip,width =0.95\columnwidth, keepaspectratio=true]{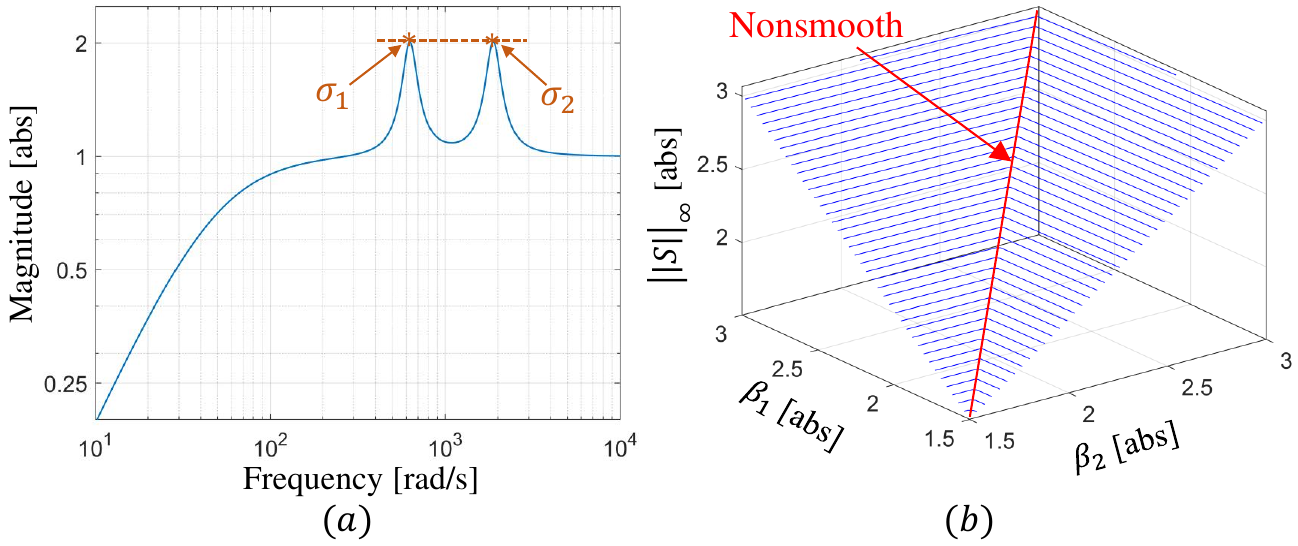}}
\vspace{-4mm}
\caption{(a) Sensitivity function. (b) Contour plot of $||S||_{\infty}$ values in terms of different $\beta_1$ and $\beta_{2}$.}
\label{fig:hinf_nonsmooth}
\vspace{-4mm}
\end{figure}

\subsubsection{Nonsmoothness in Constraint} The nonsmoothness of the constraint in \eqref{eqn:opt_form} comes from $S(j\omega)$ reaching its $H_\infty$ norm at multiple frequencies. Consider a dummy sensitivity function consisting of a high-pass filter and two peak filters as illustrated in Fig.~\ref{fig:hinf_nonsmooth}a, and $\beta_1$ and $\beta_2$ are the two parameters determining the height of each peak. The value of $||S||_{\infty}$ is determined by the higher peak. When $\beta_1 = \beta_2$, $||S||_{\infty} = S_{max}$ is attained at both peak frequencies as shown in Fig.~\ref{fig:hinf_nonsmooth}a. 
Fig.~\ref{fig:hinf_nonsmooth}b shows the contour plot of $||S||_{\infty}$ with $\beta_1$ and $\beta_2$, and nonsmoothness can be observed at the ridge $\beta_1=\beta_2$ as highlighted by the red line. At the nonsmooth locations, the steepest descent direction of $||S||_{\infty}$ is along the ridge, where a gradient cannot be obtained. 


Although the two examples illustrating the nonsmoothness are trivial, in a practical system, the nonsmoothness in  \eqref{eqn:opt_form} can become complicated when the two sources of nonsmoothness are combined and the number of controller variables increases. Regular gradient-based optimization solvers designed for smooth functions are slow and can fail as they often converge to nonstationary points \cite{asl2020analysis}, and the inclusion of parameter constraints further increases the difficulty of solving. An optimization solver tailored for constrained nonsmooth problems must be employed to effectively solve \eqref{eqn:opt_form}. 


\subsection{Nonsmooth Optimization and Solver Selection}

This section first briefly presents the fundamentals of nonsmooth optimization to make this paper self-contained. Readers are referred to \cite{clarke1990optimization} for more details. 

For a locally Lipschitz function $f: \mathbb{R}^n \rightarrow \mathbb{R}$ that is differentiable almost everywhere, its \textit{Clarke subdifferential} $\partial f(x)$ at a point $x \in \mathbb{R}^n$ is defined as 
\begin{equation}
    \begin{aligned} \label{eqn:clarke}
    \partial f(x):= \mathrm{conv} \{ \lim_{i\rightarrow \infty}&  \nabla f(x_i): x_i \rightarrow x~\\
    &\mathrm{and}~ \nabla f(x_i)~\mathrm{exists}\}.
\end{aligned}
\end{equation}
Here $\mathrm{conv}$ denotes the convex hull of a set. Each vector $g \in \partial f(x)$ is called a \textit{subgradient}. If the function $f$ is continuously differentiable at $x$, then the subdifferential reduces to a singleton as $\partial f(x)=\{ \nabla f(x)\}$.

A point $x^*$ is \textit{Clarke stationary} for $f(x)$ iff $0 \in  \partial f(x^*)$. A more robust and practical sense of stationarity of $f(x)$ can be defined by considering the limits of gradients of points near $x$. For each $\varepsilon>0$, the \textit{Clarke $\varepsilon-$subdifferential} can be defined as \cite{goldstein1977optimization}:
\begin{align}
    \partial_{\varepsilon}f(x):=\mathrm{conv}\{ \partial f(y),~ ||x-y||\leq\varepsilon   \}.
\end{align}
A point $x^*$ is \textit{Clarke $\varepsilon-$stationary} for $f(x)$ iff $0 \in  \partial_{\varepsilon} f(x^*)$. In each optimization iteration, using a single subgradient as a search direction may not be a steep descent direction, and sometimes may not even be a descent direction as shown in examples in Figs.~\ref{fig:obj_nonsmooth}-\ref{fig:hinf_nonsmooth}. To ensure effective optimization, the steepest descent direction should be computed as the minimal-norm element in $\partial_{\varepsilon}f(x)$ to accelerate convergence.

Several solvers for constrained nonsmooth problems have been developed in prior studies, including Sequential Fixed Penalty Parameter (SFPP)~\cite{gumussoy2009multiobjective}, Sequential Quadratic Programming Gradient Sampling (SQP-GS)~\cite{curtis2012sequential}, Sparse Nonlinear OPTimizer (SNOPT)~\cite{gill2005snopt}, and GRadient-based Algorithm for Non-Smooth Optimization (GRANSO)~\cite{curtis2017bfgs}. A comparison between these solvers on 200 test problems has been performed in~\cite{curtis2017bfgs}, which shows that SFPP and SNOPT could not find most of the minimizes. Among the two effective solvers (GRANSO and SQP-GS), SQP-GS is successful in more cases; however, it uses a gradient sampling method that evaluates multiple points in each iteration for descent direction search \cite{burke2020gradient} and suffers from a higher computational time.

To provide a balanced performance between optimization performance and computational efficiency, we selected GRANSO (v1.6.4 in MATLAB)  to solve  the constrained nonsmooth optimization in the bandwidth optimal control problem. GRANSO uses a BFGS-SQP  algorithm \cite{curtis2017bfgs} that employs BFGS quasi-Newton Hessian approximation in an SQP-based steering strategy to compute the search directions, and uses an inexact Armijo-Wolfe line search to update the iterations.  
To determine the convergence in a nonsmooth situation, an extra QP problem is formed with historic gradients of the objective and constraints to approximate $\partial_{\varepsilon}f(x)$. The optimization terminates if the optimum solution from the QP is smaller than a fixed tolerance and the constraints are satisfied. Although the BFGS-SQP does not have theoretical convergence guarantees, it performed well in solving challenging test problems in practice~\cite{curtis2017bfgs}.  In addition, GRANSO does not assume a special structure of the objective function and constraints, making it suitable for general nonsmooth problems, including our bandwidth optimal control \eqref{eqn:opt_form}. 


\subsection{Bandwidth Optimal Control Problem Solving}

Solving the bandwidth optimal control problem \eqref{eqn:opt_form} using the GRANSO requires computing the descent directions of the objective and the constraints in each iteration. This section presents the closed-form computation of the descent direction of the objective function ($\omega_{bw}$) and robustness constraint ($\|S\|_{\infty}$). A proposed steepest descent direction computation method aims to speed up the convergence exploiting properties of  $\omega_{bw}$ and $\|S\|_{\infty}$ is also discussed. 

\subsubsection{Descrent Direction of Objective Function}
To align with the minimizing convention in GRANSO, we first change the objective ``$\mathrm{max}$ $\omega_{bw}$'' in \eqref{eqn:opt_form} to ``$\mathrm{min}$ $-\omega_{bw}$''. 
Note that the bandwidth is not differentiable with respect to $\theta_c$ when the multiplicity of $\underline{\sigma}(L(j\omega, \theta_c)|_{\omega = \omega_{bw}})$ (where \eqref{eqn:bw_def} is attained) is greater than 1. 
Define  $\sigma_l =\underline{\sigma}(L(j\omega_{bw}, \theta_c))$ with a multiplicity of $k$, where $l=1,...,k$. 
The subgradients $g_l$ of $\omega_{bw}$ with respect to $\theta_c$ can be derived by applying the implicit function theorem 
to \eqref{eqn:bw_def} as
\begin{align} \label{eqn:L_sv_g}
    g_l = -\frac{\partial\sigma_l(L(j\omega_{bw},\theta_c))}{\partial \theta_c}\Big/ \frac{\partial\sigma_l(L(j\omega_{bw},\theta_c))}{\partial \omega_{bw}}.
\end{align} 

The derivative of an arbitrary singular value $\sigma_i$ of a general complex matrix $A$ w.r.t. a real parameter $p \in \mathbb{R}$ can be found as 
\begin{align} \label{eqn:sigma_A_p}
    \frac{\partial \sigma_i}{\partial p} = \mathrm{Real}\big[u_i^* \frac{\partial A}{\partial p}v_i\big],
\end{align}
where $u_i$ and $v_i$ are the $i$-th column of the unitary matrices $U$ and $V$ from the singular value decomposition (SVD), i.e., $A = U\Sigma V^*$. Applying \eqref{eqn:sigma_A_p} to \eqref{eqn:L_sv_g}, we have 
\begin{align}
    \frac{\partial\sigma_l}{\partial \theta_{ci}}&= \mathrm{Real} \big[ u_l^*G(j\omega_{bw}) \frac{\partial C(j\omega_{bw},\theta_c)}{\partial \theta_{cj}}  v_l  \big],
    \\
     \frac{\partial\sigma_l}{\partial \omega_{bw}}&= \mathrm{Real} \big[ u_l^* \frac{\partial L(j\omega_{bw},\theta_c)}{\partial \omega_{bw}}  v_l  \big],\label{eq:sigma_l_bw}
\end{align}
where $\theta_{ci}$ is the $i$-th component of $\theta_c$, $\frac{\partial C}{\partial \theta_{ci}}$ and $\frac{\partial L}{\partial \omega_{bw}}$ can be found once the controller structure $C$ is given, and $\frac{\partial\sigma_l}{\partial \theta_{c}} = [\frac{\partial\sigma_l}{\partial \theta_{c1}},\dots,\frac{\partial\sigma_l}{\partial \theta_{cn}}]^\top $ and $u_l,v_l$ are from the SVD of $L(j\omega_{bw},\theta_c)$. The subgradients for the objective function can be computed via \eqref{eqn:L_sv_g}-\eqref{eq:sigma_l_bw}.

The above-computed subgradients can already be used as the input for the GRANSO solver. However, in our formulation \eqref{eqn:opt_form}, the structures of objective and constraint are available, which provide valuable information to accelerate the convergence. 
Augmenting the subgradient computation mentioned above, we further propose the following method to solve for the steepest descent direction among the subdifferentials.
The subdifferentials of $\omega_{bw}$ at $\theta_c$ can be assembled as 
\begin{align} \label{eqn:sub_obj}
    \partial \omega_{bw}(\theta_c)=\mathrm{conv}\{g_l, l = 1,\dots,k\}.    
\end{align}
The steepest descent direction $d^{bw}$ can be obtained by solving the following QP problem:  
\begin{small}
    \begin{align} \label{eqn:qp_obj}
\begin{split}
    \min_{\lambda_1,\dots,\lambda_k \in \mathbb{R}_+}~~||d^{bw} = \sum_{l=1}^k \lambda_l g_l||_2~~~
    \mathrm{s.t.}~~ \sum_{l=1}^k \lambda_l = 1,
\end{split}
\end{align}
\end{small}

\noindent where $\lambda_1,\dots,\lambda_k$ are positive real numbers. The resultant $d$ is the minimum-norm element in \eqref{eqn:sub_obj}. The problem \eqref{eqn:qp_obj} is strongly convex and can be solved easily by existing solvers such as CVX or MOSEK. 

\rem{In practice, it is common that some computed singular values of $L(j\omega_{bw}, \theta_c)$ are distinct numerically but have values very close to each other, i.e., are not ``well-separated''. Under this situation, the descent direction $d$ computed from \eqref{eqn:qp_obj} can be conservative. To avoid this challenge,  the term ``singular values of multiplicity $k$'' should be modified as ``a cluster of $k$ poorly separated singular values that are well separated from all other singular values'' in practice \cite{freudenberg1982robustness}. A positive user-defined tolerance $\delta_{bw}$ can be used to determine the clustering threshold of the $\sigma_l$'s as
\begin{align}
    \sigma_l \in \{ \sigma | \sigma\leq (1+\delta_{bw})\underline{\sigma}\}.\label{eqn:tolerance}
\end{align}

\begin{figure*}[t!]
\centering
\subfloat{
\includegraphics[trim={0mm 0mm 0mm 0mm},clip,width =1.9\columnwidth, keepaspectratio=true]{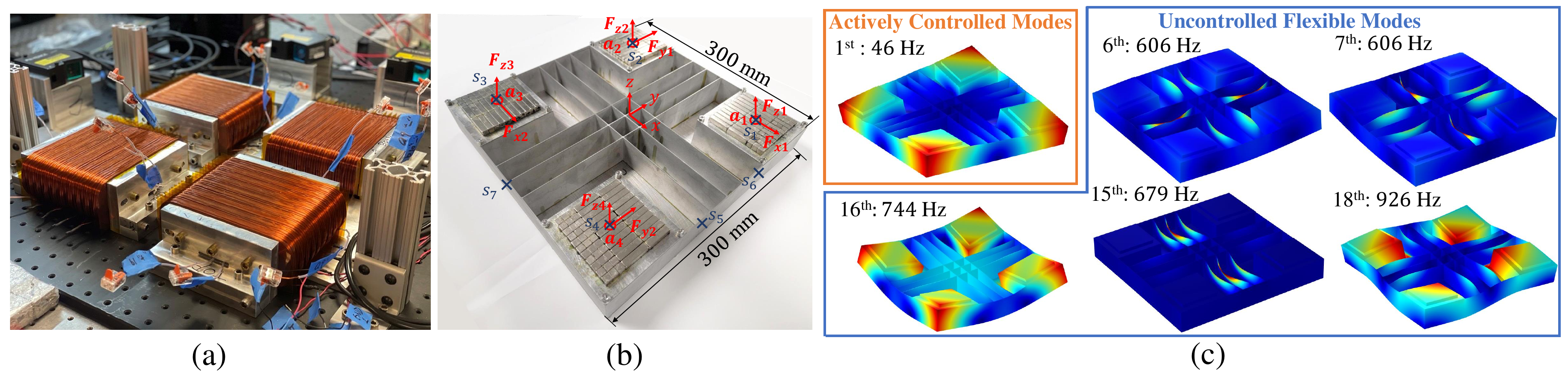}}
\vspace{-4mm}
\caption{Hardware overview of the maglev motor system for the case study. (a) Stator with coils. (b) Moving Stage with Halbach arrays. (c) Stage's flexible mode shapes that are observable and controllable. } 
\vspace{-4mm}
\label{fig:Structure}
\end{figure*}

\subsubsection{Descent Direction of Constraint}

The $H_{\infty}$-norm of the closed-loop MIMO sensitivity $||S(j\omega))||_{\infty}$ is always attained at a finite set of frequencies  $\Omega(\theta_c) =\{\omega_1,\dots,\omega_r\}$. Assume the multiplicity of $\Bar{\sigma}(S(j\omega))$ is  $k_i$ at each $\omega_i \in \Omega(\theta_c)$, $i = 1, \dots, r$. The subgradients of $||S(j\omega_i)||_{\infty}$ with respect to $\theta_c$ is \cite{giesy1993h} 
\begin{align}\label{eqn:sub_S}
    h_{il} = \mathrm{Real}  \big[{u}^{S*}_{il} \frac{\partial S(j\omega_i,\theta_c)}{\partial \theta_c} {v}^S_{il}\big],
\end{align}
where ${u}^S_{il}$ and ${v}^S_{il}$ are the singular vectors corresponding to the $l$-th singular value of $S(j\omega_i,\theta_c)$, $l = 1,\dots,k_i$. Consider the identity $\frac{\partial U^{-1}}{\partial x} = - U^{-1} \frac{\partial U}{\partial x} U^{-1}$, we have 
\begin{align}
    \frac{\partial S(j\omega_i,\theta_c)}{\partial \theta_c}=-SG(j\omega_i)\frac{\partial C(j\omega_i,\theta_c)}{\partial \theta_c}S.
\end{align}
The subdifferentials of $||S(\theta_c)||_{\infty}$ at $\theta_c$ is formed as 
\begin{align}
    \partial f_{\infty}(\theta_c) = \mathrm{conv}\{ h_{il},~\forall i,l \}.
\end{align}
The steepest descent direction ${d}^S$ can be found as the minimum-norm component in  $\partial f_{\infty}(\theta_c)$ as
\begin{small}
\begin{align} \label{eqn:qp_constraint}
\begin{split}
    \min_{\lambda \in \mathbb{R}_+^t}~||{d}^S = \sum_{i=1}^r \sum_{l=1}^{k_j} \lambda_{il} h_{il}||_2, ~~
    \mathrm{s.t.}~ \sum_{i=1}^r \sum_{l=1}^{k_i} \lambda_{il} = 1,
\end{split}
\end{align}
\end{small}

\noindent where $\lambda$ is a positive vector of dimension $t=\sum_{i=1}^r k_i.$   In practice, $\Bar{\sigma}(S(j\omega))$  is often attained at only one frequency $\omega_i \in \Omega(\theta_c)$, which significantly simplifies the QP problem \eqref{eqn:qp_constraint}. The result of \eqref{eqn:qp_constraint} ${d}^S$ provides the steepest descent direction of the robust constraint and can serve as an input of the GRANSO solver.

\rem{For a practical system, it is unlikely that the sensitivity function $S(j\omega)$ can attain the same  $H_{\infty}$ norm at multiple frequencies. However, it is highly likely that $S$ has multiple peak magnitudes that are sufficiently close to the $H_{\infty}$ norm, and the frequencies of these peaks can be collected into $\Omega(\theta_c)$. The threshold can be controlled by a user-defined tolerance~$\delta_h$ in a similar manner with \eqref{eqn:tolerance}. }

\subsection{Practical Issues}

\subsubsection{Initization}
To start the solving iterations, the initial controller must achieve closed-loop stability to attain a finite value in $||S||_{\infty}$. One approach to reach a stable controller is formulating a stabilization optimization as introduced in \cite{apkarian2006nonsmooth} for a general controller structure. In this work, with the dynamics of the plant system available, the initial controller can be synthesized via manual tuning. 
In addition, the GRANSO solver only provides local optimality, and multiple randomized initializations are needed to capture the global optimum. In this case, synthesizing initial stable controllers is of significant importance. Due to the length limit, the global version of the bandwidth optimal control is not presented in this paper, and will be studied in future work. 

\subsubsection{Hyperparameter Tuning}
The GRANSO solver has several hyperparameters and their tuning is critical to the success of optimization. For example, the steering parameters $c_v$ and $c_{\mu}$ determine the extent of promoting progress towards feasibility. For the bandwidth optimal control problem  \eqref{eqn:opt_form},  if the cross-coupling effect is severe between different decoupled channels, an aggressive set of $c_v$ and $c_{\mu}$ is necessary to enforce robustness. Conversely, for systems with insignificant challenges to the robustness constraint, selecting a set of overly aggressive steering parameters can lead to slow convergence. The proper values can usually be tuned by observing the plant dynamics and running trial iterations. Another important hyperparameter is the initial penalty parameter $\mu$, which is assigned to the objective value in the penalty function. Due to the nature of the steering strategy, the penalty parameter decreases during the iterations. As a result, if the magnitude of the objective function is significantly smaller than that of the constraints, a large initial value should be selected for $\mu$.

\subsubsection{Parameter Scaling}
The controller parameters in $\theta_c$ can take a large range in their values since they have different units, which can lead to an ill-conditioned Hessian and thus failure in optimization \cite{li1995mechanical}. In practical solving, proper scaling is required among all the decision variables to make them of the same order of magnitude. 



\section{Simulation Evaluation}\label{sec:simulations}


\subsection{Case Study Motion System Overview}

\begin{figure}[t!]
\centering
\subfloat{
\includegraphics[trim={0mm 0mm 0mm 0mm},clip,width =1\columnwidth, keepaspectratio=true]{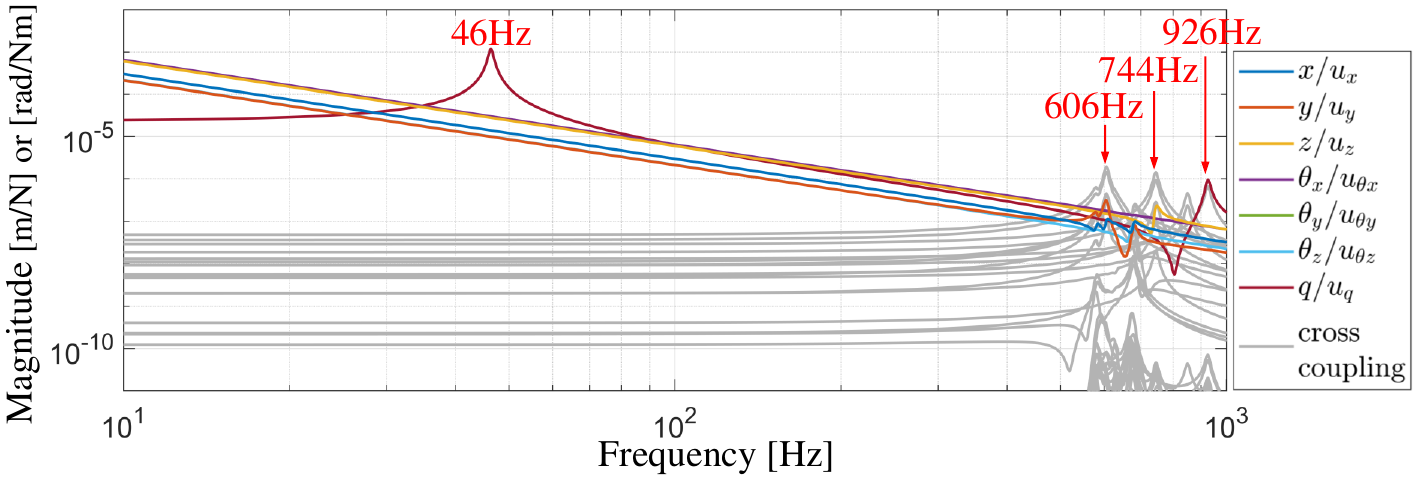}}
\vspace{-4mm}
\caption{Frequency response of the plant motion system in decoupled DOFs and cross-coupling terms.} 
\vspace{-4mm}
\label{fig:plant_tf}
\end{figure}

Fig.~\ref{fig:Structure}a-b shows the photos of the FleXstage system \cite{wu2023flexstage}, which is being used as a case study in this paper. The moving stage is 300~mm~$\times$~300~mm in size. The system's magnetic design follows Kim et al.~\cite{kim1998high}, where four permanent magnet arrays are located at the corners of the stage to provide both vertical levitation forces and lateral thrust forces, as shown by red arrows in Fig.~\ref{fig:Structure}b.  The lightweight stage structure uses ribs to reinforce a thin stage top. To overcome the trade-off between achievable lightweight and control bandwidth, in this design, the stage’s first flexible mode is intentionally designed to be compliant (resonance 50~Hz, well within the target control bandwidth), and the rest of the excitable flexible modes are stiffened to have resonance
frequencies above 600~Hz (Fig.~\ref{fig:Structure}c). Feedback control is conducted for six rigid-body DOFs to achieve magnetic levitation; in addition, the first flexible mode is also actively controlled with a bandwidth beyond its resonance frequency to introduce ``servo stiffness''~\cite{wu2022sequential}. Seven displacement sensors are used to measure the stage's position and deformation, with the sensor locations labeled via blue crosses in Fig.~\ref{fig:Structure}b. 

Define the vector of eight actuation force  as the control input $u$, and the vector of seven sensor signals as the measurement $y$. The system dynamics in the form of \eqref{eqn:mech_EOM} is obtained from finite element simulation (using COMSOL Multiphysics). This dynamic model is then decoupled into seven SISO channels as described in Section~\ref{sec:formulation}.  Fig.~\ref{fig:plant_tf} shows the plant frequency responses of the seven decoupled SISO channels (including six rigid body motion DOFs and one flexible mode with modal displacement $q$) as well as the cross-coupling terms in grey lines. It can be observed that the cross-coupling terms have large magnitude at several resonance frequencies (e.g., 606~Hz and 744~Hz), which can cause robustness and even stability challenges when the system's control bandwidth increases. 

\subsection{Bandwith Optimal Control Evaluation}

The proposed nonsmooth-optimization-based bandwidth optimal control is evaluated by the controller synthesis for the FleXstage system. A decentralized controller $C=\mathrm{diag}\{ C_1,\dots, C_7\}$ is used for the decoupled channels, and each SISO controller $C_i, i = 1, \dots, 7$ follows the structure shown in \eqref{eqn:PID}. With each SISO controller having only one parameter $\omega_{ci}$, there are seven controller parameters to be determined in this problem.

\begin{figure}[t!] 
\centering
\subfloat{
\includegraphics[trim={0mm 0mm 0mm 0mm},clip,width =0.95\columnwidth, keepaspectratio=true]{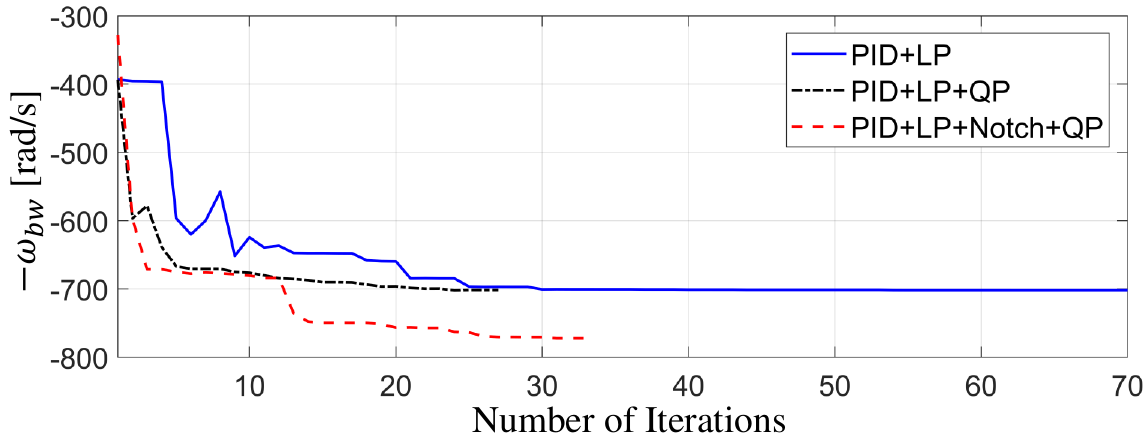}}
\vspace{-4mm}
\caption{History of convergence in for bandwidth optimal control problem for three evaluation cases. }
\label{fig:history}
\vspace{-3mm}
\end{figure}

\begin{table}[t] 
\centering
\caption{Convergence comparisons}
\vspace{-0mm}
\label{table:Convergence}
    \begin{center} \begin{small}
        \begin{tabular}{ >{\centering\arraybackslash}p{2.3cm} >{\centering\arraybackslash}p{1.25cm}  >{\centering\arraybackslash}p{0.8cm} >{\centering\arraybackslash}p{1.75cm}}
        \hline
        Test Case & Opt. Obj. & \# Iter. & \# Fun. Eval. \\
        \hline
        PID+LP  & $-701.8$  & 70  & 442
        \\
        PID+LP+QP  & $-701.6$  & 27  & 216
        \\
        PID+LP+Notch+QP  & $-772.2$  & 33  & 225\\

        \hline

        \end{tabular}
    \end{small} \end{center}
\vspace{-6mm}
\end{table}

To evaluate the effectiveness of the proposed QP-based steepest descent direction search \eqref{eqn:qp_obj} and \eqref{eqn:qp_constraint}, the bandwidth optimal control problem for the FleXstage system is solved in two different methods. The first method, which we call ``PID+LP'', directly feeds the subgradients \eqref{eqn:sub_obj} and \eqref{eqn:sub_S} into the GRANSO solver. The second method, which we call ``PID+LP+QP'', solves the QP problems  \eqref{eqn:qp_obj} and \eqref{eqn:qp_constraint} to search for the steepest descent direction in the subdifferential set and then use them as the GRANSO solver input. $S_{max}= 2$ is selected, and the steering parameters are  $c_v=0.7,c_{\mu}=0.3$. Same initial controller parameters are used for two test cases, with $\omega_{c0} = 377~\rm{rad/s}$ for all rigid-body motion control channels, and $\omega_{c0} = 439~\rm{rad/s}$ for the flexible mode control channel allow for the initial bandwidth higher than the resonance frequency. The tolerances of separated singular values of $L$ and $S$ are set to be $\delta_{bw}=0.02$ and $\delta_h=0.005$, respectively.  Fig.~\ref{fig:history} shows the history of the objective function through the optimization iterations. Data show that both tests can converge to the same feasible optimum upon termination. However, the test ``PID+LP'' takes a significantly larger number of iterations to converge, and its history of the objective demonstrations has multiple flat regions especially when close to converging. This is mainly because there exist not-well-separated singular values in both objective function and constraints, and using the subgradients as the descent direction can lead to conservative performance. 
In contrast, the ``PID+LP+QP'' test demonstrates faster and more monotonic convergence since the QP problems  \eqref{eqn:qp_obj} and \eqref{eqn:qp_constraint} can effectively identify the steepest descent direction. Table~\ref{table:Convergence} compares the optimal objective values, number of iterations,  and number of function evaluations of the two test cases. Notably, compared with the ``PID+LP'' baseline, the proposed ``PID+LP+QP'' method reduces the iterations and function evaluations by 61\% and 51\%, respectively, demonstrating the effectiveness of the QP descent direction search \eqref{eqn:qp_obj} and \eqref{eqn:qp_constraint} in facilitating convergence.

\begin{figure}[t!] 
\centering
\subfloat{
\includegraphics[trim={0mm 0mm 0mm 0mm},clip,width =0.9\columnwidth, keepaspectratio=true]{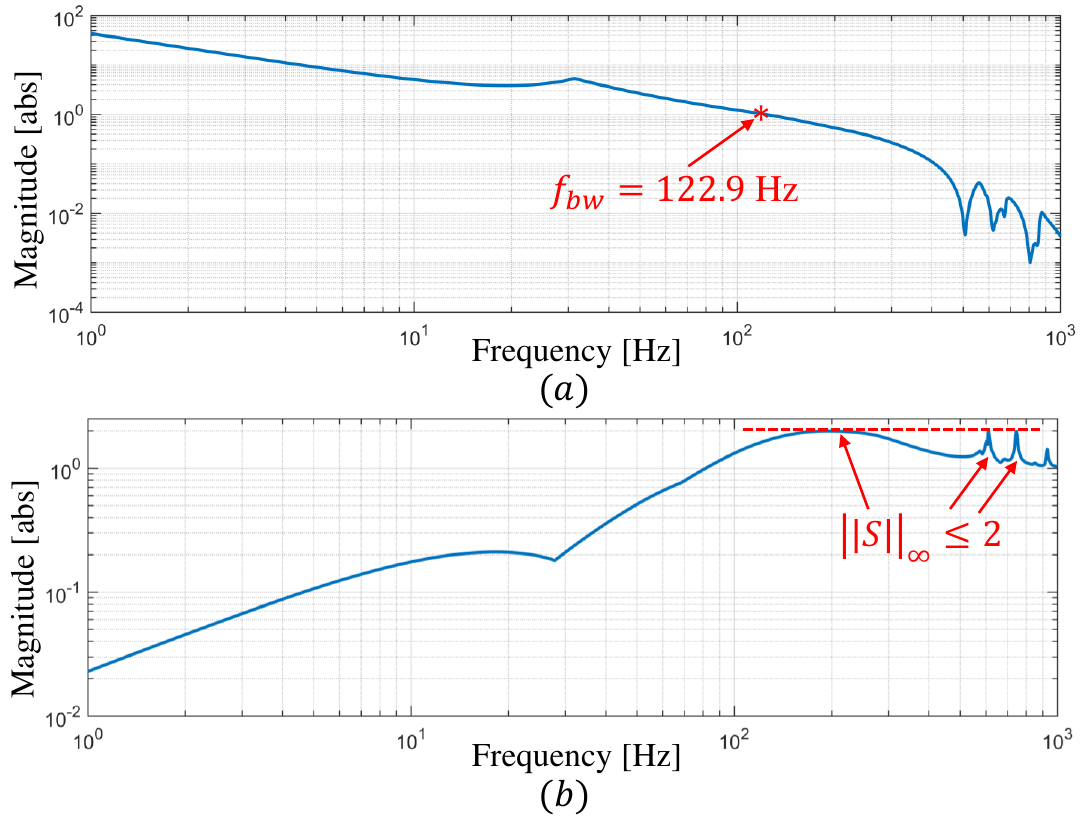}}
\vspace{-4mm}
\caption{(a) Optimal minimum singular values of Loop Gain (b) Optimal maximum singular values of Sensitivity}
\label{fig:closed_loop}
 \vspace{-6mm}
\end{figure}

Although the aforementioned test cases already demonstrate excellent performance in control bandwidth (701~rad/s or 112~Hz) while maintaining robustness, this controller design is still conservative since the selected SISO controller structure \eqref{eqn:PID} cannot effectively address the bandwidth limit due to structural resonances. To further increase the control bandwidth, the bandwidth optimal control problem for the FleXstage system is solved again with notch filters \eqref{eqn:notch} included in the controllers for channels with structural resonances limiting the control bandwidth. This test is called ``PID+LP+Notch+QP''. Based on the plant frequency response shown in Fig.~\ref{fig:plant_tf}, we decided to include notch filters in the controllers for the three translational motion channels and the yaw control channel. The notch frequencies are placed at their corresponding resonance frequencies (shown in Fig.~\ref{fig:plant_tf}), and the notch width and depth are controller parameters to be solved, which increases the number of controller parameters to 15. Note that the value of $\omega_c$'s are around 700 and the $\beta,\zeta$'s are below 1, and proper scaling of parameters must be performed to avoid numerical challenges. The QP-based steepest descent direction searches are included, $S_{max}= 2$ is selected, and the steering parameters are $c_v=0.8$ and $c_{\mu}=0.2$. Fig.~\ref{fig:history} and Table~\ref{table:Convergence} show the evaluation results. 
Compared to ``PID+LP+QP'', the system's bandwidth is increased by 10\% by introducing the notch filters with only six more iterations and nine more function evaluations. The optimal loop gain and closed-loop sensitivity of the FleXstage system solved by the ``PID+LP+Notch+QP'' are shown in Fig.~\ref{fig:closed_loop}. Data show that the synthesized controller demonstrates excellent control bandwidth (772~rad/s or 123~Hz) with a sensitivity below the constraint value 2. This bandwidth is 23\% higher than that of the manual loop shaping design reported in \cite{wu2023flexstage}, which demonstrates the effectiveness of our proposed bandwidth optimal control framework.



\section{Conclusions and Future Work}\label{sec:conclusion}

In this paper, we proposed a novel bandwidth optimal control framework based on nonsmooth optimization, aiming to efficiently synthesize controllers with high bandwidth and guaranteed robustness, which is of critical importance in motion systems. The bandwidth optical control problem is formulated and its nonsmoothness in both objective function and constraints is illustrated. Solving the problem using the GRANSO solver with descent direction calculation is discussed in detail. The proposed framework successfully synthesized a decentralized controller for a precision positioning system that has high control bandwidth and satisfies robustness constraints. 
Future work will consider (a) improving the bandwidth optimal control framework to reach the global optimum via randomized initialization, (b) adopting a general controller structure to push for better overall performance, and (c) experimentally evaluating the synthesized controllers for performance demonstration. 





\balance
\bibliographystyle{IEEEtran}
\bibliography{main}

\end{document}